\documentclass[conference]{IEEEtran}
\IEEEoverridecommandlockouts
\usepackage{cite}
\usepackage{subcaption}
\usepackage{makecell}
\usepackage{amsmath,amssymb,amsfonts}
\usepackage{algorithmic}
\usepackage{graphicx}
\usepackage{textcomp}
\usepackage{xcolor}
\usepackage{url}
\usepackage{multirow} 
\usepackage{booktabs}
\def\BibTeX{{\rm B\kern-.05em{\sc i\kern-.025em b}\kern-.08em
    T\kern-.1667em\lower.7ex\hbox{E}\kern-.125emX}}
\begin{document}

\title{Interactive Drawing Guidance for Anime Illustrations with Diffusion Model}

\author{\IEEEauthorblockN{Chuang Chen\IEEEauthorrefmark{1}, 
Xiaoxuan Xie\IEEEauthorrefmark{1}, 
Yongming Zhang, 
Tianyu Zhang, 
Haoran Xie\IEEEauthorrefmark{2}}

\IEEEauthorblockA{Japan Advanced Institute of Science and Technology}
\IEEEauthorblockA{Ishikawa, Japan}
}

\maketitle

\newcommand\blfootnote[1]{%
  \begingroup
  \renewcommand\thefootnote{}\footnote{#1}%
  \addtocounter{footnote}{-1}%
  \endgroup
}
\blfootnote{\IEEEauthorrefmark{1}Same contribution.} 
\blfootnote{\IEEEauthorrefmark{2}Corresponding author (xie@jaist.ac.jp).} 

\begin{abstract}
Creating high-quality anime illustrations presents notable challenges, particularly for beginners, due to the intricate styles and fine details inherent in anime art. We present an interactive drawing guidance system specifically designed for anime illustrations to address this issue. It offers real-time guidance to help users refine their work and streamline the creative process. Our system is built upon the StreamDiffusion pipeline to deliver real-time drawing assistance. We fine-tune Stable Diffusion with LoRA to synthesize anime style RGB images from user-provided hand-drawn sketches and prompts. Leveraging the Informative Drawings model, we transform these RGB images into rough sketches, which are further refined into structured guidance sketches using a custom-designed optimizer. The proposed system offers precise, real-time guidance aligned with the creative intent of the user, significantly enhancing both the efficiency and accuracy of the drawing process. To assess the effectiveness of our approach, we conducted a user study, gathering empirical feedback on both system performance and interface usability.

\end{abstract}


\section{Introduction}

Anime illustration is a globally influential art form with significant commercial value, widely applied in the entertainment, merchandise, and gaming industries. However, despite its seemingly simple appearance, anime drawing presents significant challenges to beginners. Achieving high-quality results requires advanced skills, including mastery of proportions, poses, and facial expressions. To address this, we propose an anime illustration drawing assistance system that helps users practice their skills and refine their creative ideas.

The drawing assistance systems have been developed to help users achieve their creative goals or improve their drawing proficiency. The early approaches used data retrieval techniques to dynamically match users' hand-drawn sketches with similar samples from a database~\cite{5959134,10.1145/2010324.1964922}. PortraitSketch \cite{10.1145/2642918.2647399} further refined user's hand-drawn sketches by allowing image uploads and line tracing, resulting in aesthetically enhanced drawings. More recently, generative models have been widely applied in drawing guidance systems \cite{ghosh2019isketchnfill}. However, these systems often depend on specialized datasets and the user's drawing ability, limiting creative flexibility within the scope of the dataset.

Diffusion models~\cite{ho2020denoising} have recently gained significant attention in image generation due to their diversity, stability, and high-quality outputs. Compared to traditional generative models, diffusion models offer greater flexibility in producing detailed and diverse images. Our system integrates four StreamDiffusion  \cite{streamdiffusion} pipelines, enabling real-time image generation and seamless interaction with text-to-image models such as Stable Diffusion \cite{rombach2022highresolutionimagesynthesislatent}, known for high-quality anime style synthesis. Additionally, we employ Low-Rank Adaptation (LoRA) \cite{hu2021lora} to fine-tune the model with a limited number of samples, facilitating efficient adaptation to anime illustrations. In addition, we incorporate the Informative Drawings model \cite{chan2022learning} along with a custom-designed optimizer to generate structured guidance sketches. Our system provides intuitive, flexible, and real-time anime illustration guidance, allowing users to express their creative intent more effectively. To evaluate its effectiveness, we conducted user studies comparing various drawing assistance interfaces, assessed drawing quality through expert and non-expert reviews, and measured usability using the System Usability Scale (SUS).

In this work, our key contributions are listed as follows:
\begin{itemize}
\item We propose a real-time anime illustration drawing guidance system that enables users to input both hand-drawn sketches and textual prompts, enhancing flexibility in creative expression.

\item We introduce a novel drawing guidance interface that allows users to adjust the position of the guidance sketches dynamically. A user study demonstrates that the proposed interface significantly helps users achieve their desired sketches with ease.

\end{itemize}

\begin{figure*}[!ht]
    \centering
    \includegraphics[width=\linewidth]{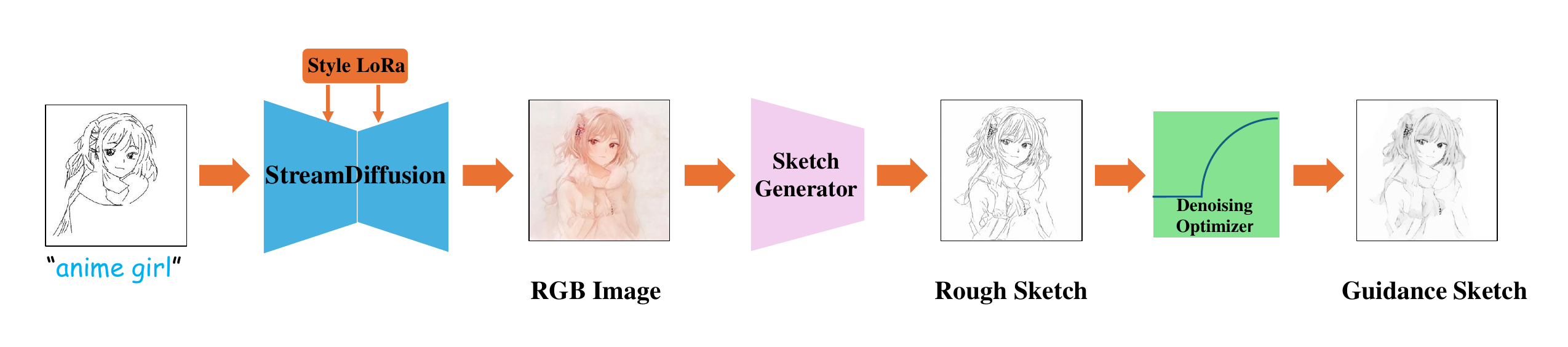}
    \caption{The pipeline of our system begins with a hand-drawn sketch and a text prompt as inputs. StreamDiffusion is utilized to generate image, which is then processed by the Sketch Generator to extract the rough sketch. The rough sketch may contain some noise, which is subsequently removed by the denoising optimizer, resulting in a clean guidance sketch for anime illustration.}
    \label{fig:framework}
\end{figure*}

\section{Related Works}

\subsection{Drawing Guidance System}

The drawing guidance system not only enables users to experience the joy of drawing and boosts their confidence, but also effectively enhances their drawing abilities. Dualface~\cite{huang2022dualface} introduces a two-stage drawing guidance interface specifically tailored for freehand portrait sketching. Sketchhelper~\cite{8607060} generates shadow guidance for users by retrieving strokes from the dataset in real time. EZ-Sketching~\cite{10.1145/2601097.2601202} automatically corrects hand-drawn sketch lines roughly traced over an image by a three-level optimization method. The features of the drawing interface with grid guidance~\cite{kana23} and variable canvas sizes~\cite{wang24} may influence the drawing results. Moreover, systems such as AniFaceDrawing~\cite{huang2023Anifacedraw} propose an unsupervised stroke-level decoupled training strategy that enables sparse sketches to automatically be matched to the corresponding localizations of anime portraits. However, these systems predominantly focus on facial drawing, thereby limiting the user's creativity. To address this limitation, we propose a system capable of drawing objects in real time. Our system significantly expands users' creative freedom and enhances interactive experiences.

\subsection{Sketch-based Applications}

Sketches can quickly and intuitively express creative ideas and design concepts, playing a crucial role in fields such as content design and engineering modeling. Sketch2Model~\cite{zhang2021sketch2model} converts hand-drawn sketches into 3D models, providing users with a convenient tool for 3D design. Additionally, Pix2Pix~\cite{isola2018imagetoimagetranslationconditionaladversarial} transforms sketches into realistic images, which are widely utilized across various design and creative disciplines. Moreover, Sketch-RNN\cite{ha2017neuralrepresentationsketchdrawings} introduces machine learning techniques to automatically and intelligently generate sketches. In summary, these methods not only enhance the efficiency of sketch utilization in the design process but also broaden their application scope, positioning sketches as pivotal tools in numerous fields.

\subsection{Diffusion Models}

Traditional retrieval-based methods rely heavily on the target databases and struggle to provide effective guidance when user input falls outside the scope of existing data, significantly constraining the design freedom~\cite{su2014ez,xie2014portraitsketch}. Bhunia et al.~\cite{bhunia2022sketching} further observed that in real‑time drawing scenarios, retrieval performance does not improve monotonically as sketches evolve and may even decline during the drawing process. In contrast to these retrieval-based methods, diffusion models are not limited by preexisting image collections and can synthesize entirely novel styles and shapes, thereby more effectively supporting diverse user creativity. Stable Diffusion models~\cite{rombach2022highresolutionimagesynthesislatent} can achieve image generation with high fidelity and adaptability. ControlNet~\cite{zhang2023addingconditionalcontroltexttoimage} can enhance latent diffusion model by introducing spatial conditioning, overcoming layout constraints, and enabling users to create complex and detailed images. StreamDiffusion~\cite{streamdiffusion} achieved real‑time, efficient image generation through optimized computational workflows and parallel processing. LoRA~\cite{hu2021lora} adopted a lightweight network structure and efficient training methods, and can extend the practicality and application scope of diffusion models. In this work, we aim to propose sketch‑based drawing guidance system to deliver an intuitive and higher‑quality creative experience.

\section{Proposed Method}

Our system generates anime illustration drawing guidance sketches in real-time. We first introduce the components of the interface (Section~\ref{section:UI}). Figure \ref{fig:framework} illustrates the system pipeline, comprising three main components: StreamDiffusion, sketch generator, and denoising optimizer. Users’ hand-drawn sketches and prompts are input into StreamDiffusion to generate an RGB image (Section \ref{section:IG}). Anime Illustration LoRA is applied to aid RGB image generation (Section \ref{section:style}). The sketch generator then processes the RGB image into a rough sketch, which is further refined by our custom-designed optimization (Section \ref{section:SG}).

\subsection{User Interface}
\label{section:UI}

\begin{figure}[t]
    \centering
    \includegraphics[width=0.47\textwidth]{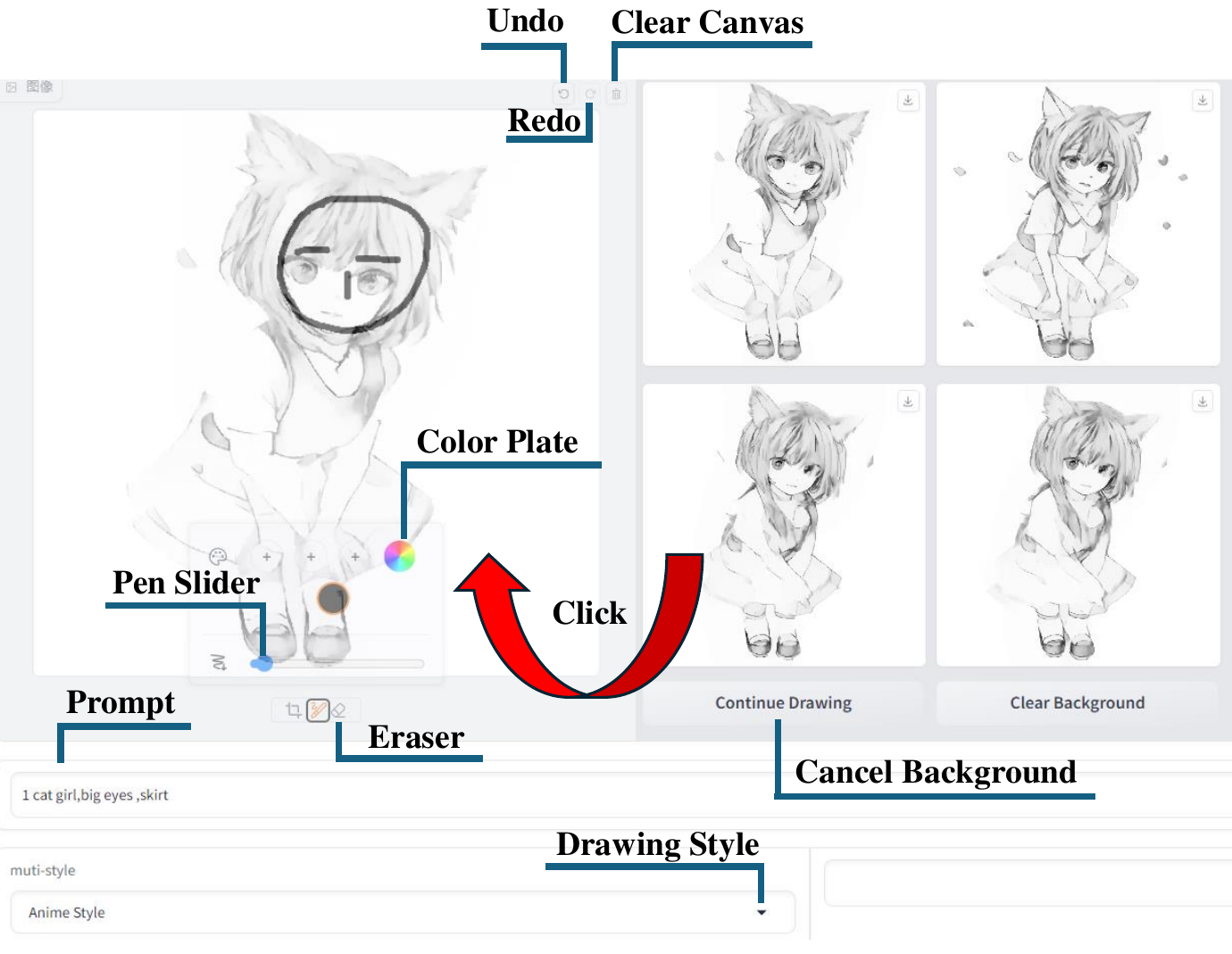}
    \caption{Interface includes essential drawing tools, enabling users to freely set, continue generating, 
    or clear background guidance sketches as needed and specify their intent through prompts and style choices.}
    \label{fig:interface}
\end{figure}

Our interface aims to assist users in drawing anime illustrations. For users who struggle to effectively convey their ideas, we provide the prompt feature. The prompt feature allows users to describe their intended sketches, generating sketches tailored to their needs. In addition, users can choose ``anime style" and ``realistic style" from a predefined list. We also offer a range of tools designed to enhance the drawing process, as shown in Figure \ref{fig:interface}.

In contrast to traditional drawing assistance systems, our method allows users to decide whether to place the anime illustration drawing guidance sketch on the background layer of the canvas to aid their drawing. Capturing users' intentions is a challenging task. If the generated content does not meet their expectations, it can become a visual distraction. Therefore, we position anime illustration drawing guidance sketches on the right side of the canvas, allowing users to reference them for inspiration without resorting to mere tracing. If users struggle to draw what they envision, they can choose to fix a preferred guidance sketch on the background layer for tracing, helping to improve their drawing skills. A new reference image is generated only after each stroke ends, avoiding frequent interruptions during the drawing process. When the user clicks one of the guidance sketch images, the system will enter a pause state, placing the selected guidance sketch on the background layer of the canvas to assist drawing. If the user clicks the “Clear Background” button, the background reference image is removed, and the system remains paused until the user explicitly clicks “Continue Drawing” to resume generation.

\subsection{Image Generation}
\label{section:IG}
Real-time performance is a critical factor affecting the interactive experience in the guidance interfaces. The StreamDiffusion \cite{streamdiffusion} inference pipeline can generate RGB images in real-time. During the hand-drawing process, we skip generating guidance sketches for extremely rough hand-drawn images to avoid wasting computational resources, such as when users repeatedly sketch the same brush stroke. The probability of skipping the image generation step is calculated as follows:

\begin{equation} 
P(\text{skip} | x) = \max\left(0, \frac{x -  \tau}{1 -  \tau}\right) 
\end{equation}

\noindent
where \( x \) represents the cosine similarity between the hand-drawn sketch \( I_i \) and the guidance sketch \( I_{\text{ref}} \) as follows:

\begin{equation} 
x = \frac{I_i \cdot I_{\text{ref}}}{\lVert I_i \rVert \lVert I_{\text{ref}} \rVert}
\end{equation}

\noindent
where \( \lVert I_i \rVert \) and \( \lVert I_{\text{ref}} \rVert \) denote their respective magnitudes. The threshold \( \tau \) controls the skipping behavior, ensuring that redundant computations are avoided when the similarity is high. Regardless of whether the image generation process is skipped, the current input image is defined as the new guidance sketch. Additionally, when the prompt changes, the system regenerates the RGB image to align with the user's requirements, independent of the similarity between sketches.

We adopt the user-drawn sketch image as the RGB input to the system, which is first encoded into a latent tensor using a Variational Autoencoder (VAE) \cite{kingma2022autoencodingvariationalbayes} and then placed into the input queue. This latent is subsequently fed into the U-Net~\cite{ronneberger2015unetconvolutionalnetworksbiomedical}, where it first passes through convolutional layers for feature extraction, followed by multiple downsampling layers, residual blocks, and upsampling modules to perform noise prediction at the designated timesteps. The entire denoising process is guided by a scheduler, which updates the latent state at each step based on the current timestep, gradually refining it into a clean image. During this process, the U-Net network handles latent tensors in parallel batches. The final latent output is decoded by the VAE into an RGB image as the final output. To further enhance performance, we adopted StreamDiffusion~\cite{streamdiffusion}, which introduced caching mechanisms, including Noise Cache, Scheduler Cache, and Prompt Embed Cache, to reduce redundant computations and accelerate the generation process. At a resolution of 512×512, generating four images in a single round takes approximately 1.2 seconds, with a memory usage of 13,244 MiB. Inference is performed using float16 precision, and the LCM-LoRA~\cite{luo2023lcm} model is fused to accelerate processing.
To prevent over-assistance, the proposed system allows control over the generation behavior by adjusting the classifier‑free guidance (CFG)~\cite{ho2022classifier} mode and guidance scale. With a lower guidance scale, the output closely follows the user’s input, generating only the drawn parts; with a higher guidance scale, the system can complete a full image even from minimal strokes, enabling flexible control from precise adherence to creative completion.

\subsection{Anime Style Generation}
\label{section:style}

\begin{figure}[t]
    \centering
    \includegraphics[width=\linewidth]{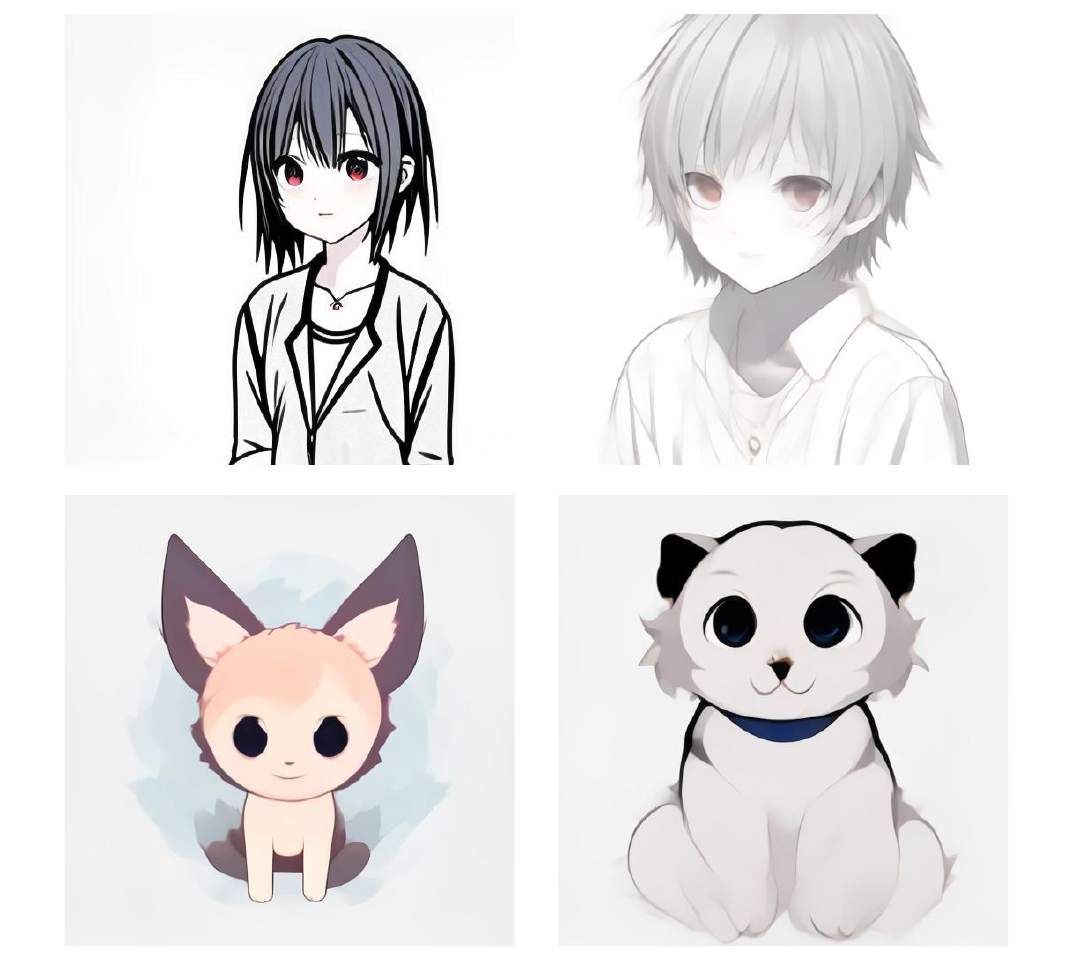}
    \caption{The four RGB images are generated by our fine-tuned model using LoRA, while the remaining images are user-drawn sketches based on these four references. }

    \label{fig:anime}
\end{figure}

To provide users with anime illustration drawing guidance, we employ LoRA \cite{hu2021lora} to fine‑tune the Stable Diffusion model to the desired style. One key advantage of LoRA is its efficiency: it adapts the model to the desired style with only a few dozen images, making it suitable for our work. We constructed a dataset of 40 high‑quality anime images, of which 17 were sourced from Pixiv\footnote{https://www.pixiv.net/} and 23 from PixAi\footnote{https://pixai.art/}. This dataset comprises 26 human characters, encompassing diverse genders and poses, and 14 animal figures, including both standard anime animals and anthropomorphic designs such as kemonomimi girls and pet‑like humanoids. The images cover various categories and styles, ranging from everyday and fantasy motifs to classic anime attire such as school uniforms, casual wear, and battle outfits. During sample selection, we prioritized clear, high‑resolution images exhibiting hallmark anime features, such as large eyes and exaggerated facial expressions, and manually annotated each sample with detailed prompts for LoRA training. Throughout fine‑tuning, the base diffusion model weights remained frozen, with updates applied exclusively to the LoRA layers. Once training was complete, the fine‑tuned LoRA parameters were seamlessly integrated into our system, providing users with precise anime illustration assistance through an intuitive interface. Figure \ref{fig:anime} showcases images generated by our drawing assistance system using the anime LoRA, which was fine-tuned on 40 anime images.

\subsection{Sketch Generation and Optimization}
\label{section:SG}
We use the Informative Drawings model~\cite{chan2022learning} to convert images into sketches. The process begins by passing the RGB image through an encoder-decoder architecture with several ResNet \cite{He_2016_CVPR} blocks to extract and process features. The initial line drawing is then subjected to depth prediction to compute geometric loss, ensuring accurate delineation of geometrically important areas. Simultaneously, the line drawing is processed with CLIP \cite{radford2021learning} to create embeddings, which are compared with the CLIP embeddings of the original image to compute semantic loss. This ensures that the generated sketch effectively conveys the semantic content of the original image. This approach allows for the efficient conversion of images into sketches.

To further refine the generated sketches, we have designed an optimizer that employs a recursive filter, an efficient image-smoothing technique. This filter achieves two-dimensional smoothing by repeatedly applying a one-dimensional filter. This approach effectively removes noise and unwanted details while preserving the image's edge information. As a result, the guidance sketches presented to users are smoother and more refined, enhancing the overall drawing experience.

\begin{figure}[ht]
    \centering
    \begin{subfigure}[b]{0.45\linewidth}
        \centering
        \includegraphics[width=\linewidth]{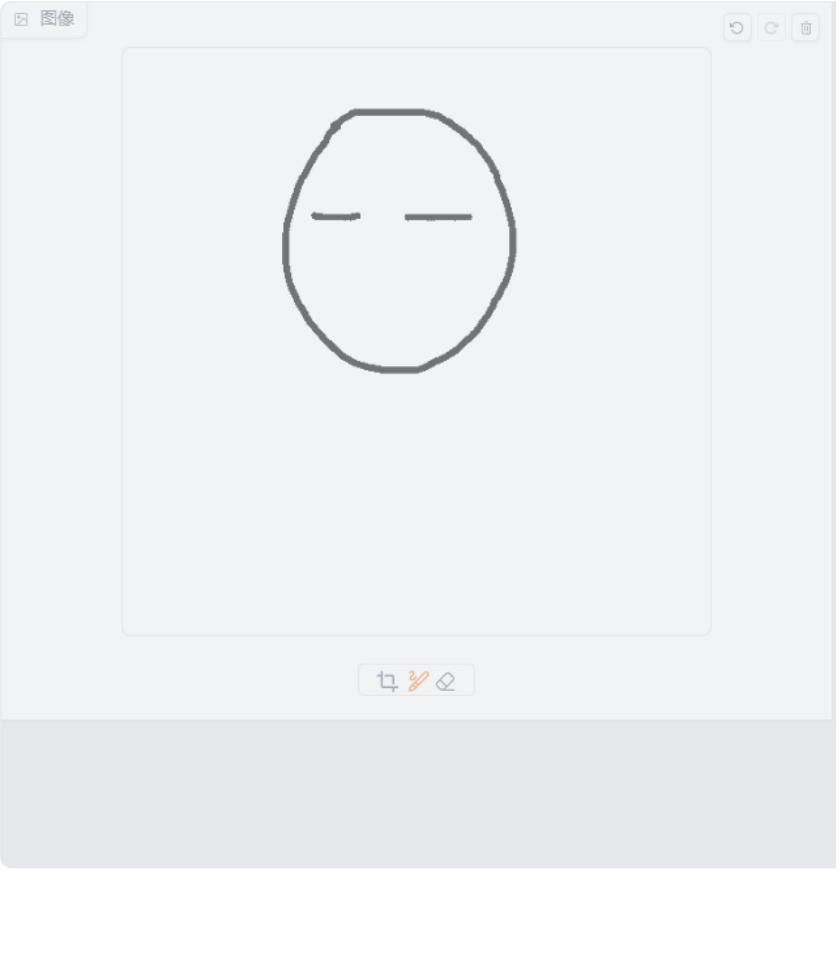}
        \caption{}
        \label{fig:ui(a)}
    \end{subfigure}
    \hfill
    \begin{subfigure}[b]{0.45\linewidth}
        \centering
        \includegraphics[width=\linewidth]{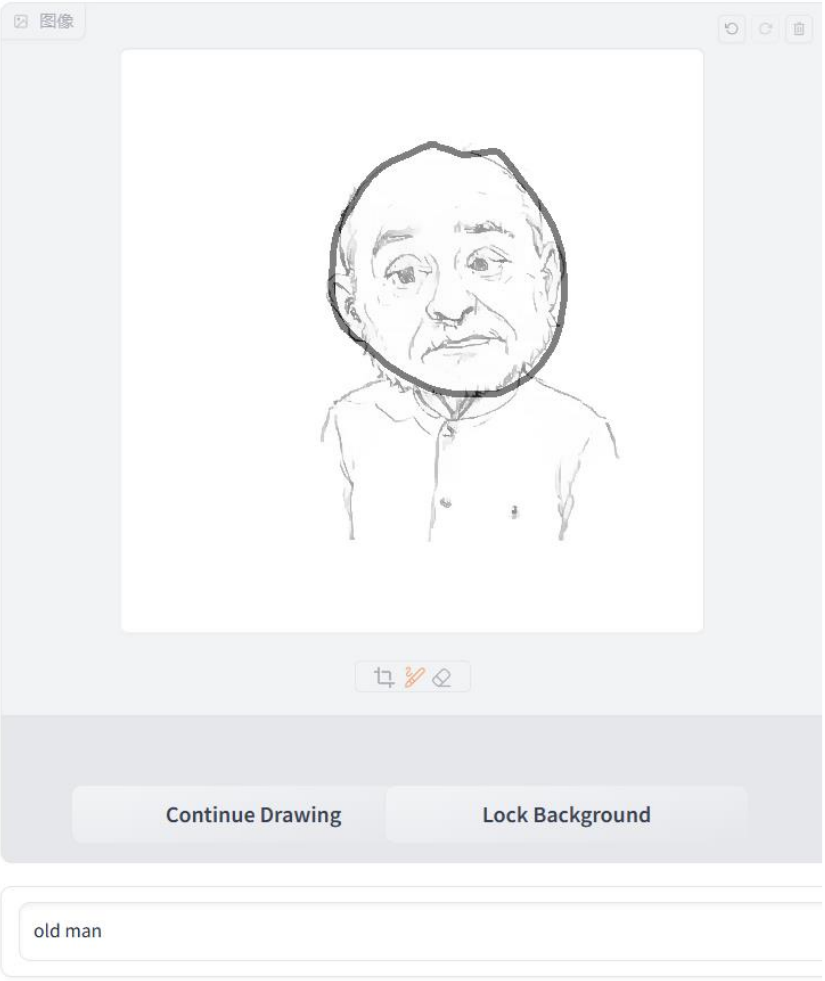}
        \caption{}
        \label{fig:ui(b)}
    \end{subfigure}
    \caption{Drawing interfaces used in our study. (a) Baseline interface: a canvas without user guidance; (b) Shadow guidance interface: interface with one guidance sketch places under the canvas.}

    \label{fig:ui}
\end{figure}

\begin{figure*}[t]
    \centering
    \includegraphics[width=0.95\linewidth]{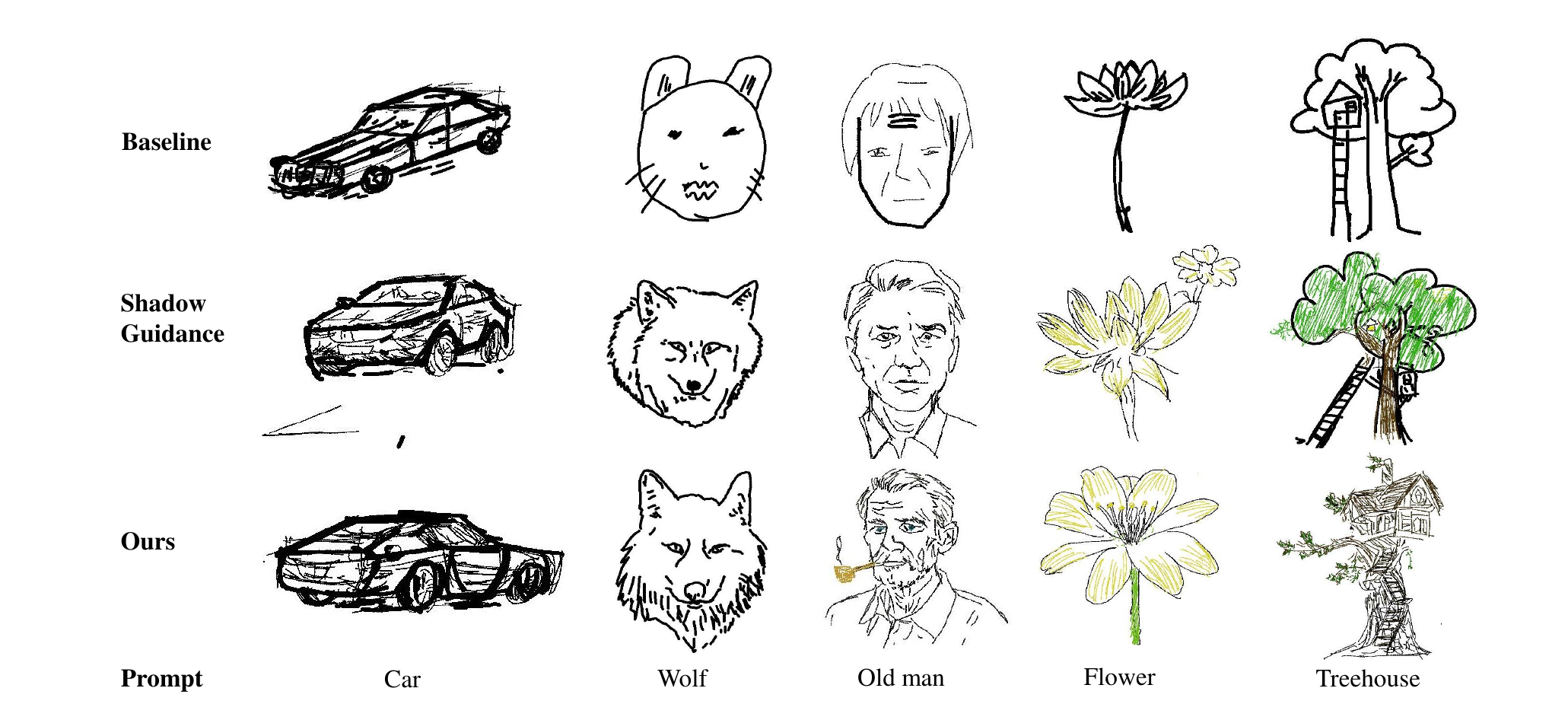}
    \caption{User study results are presented. The first column shows drawings from a professional user, while the remaining columns display results from users without drawing experience. These drawings were generated by selecting the ``realistic style" from the system's predefined list of styles.}

    \label{fig:results1}
\end{figure*}

\begin{figure*}[t]
    \centering
    \includegraphics[width=0.95\linewidth]{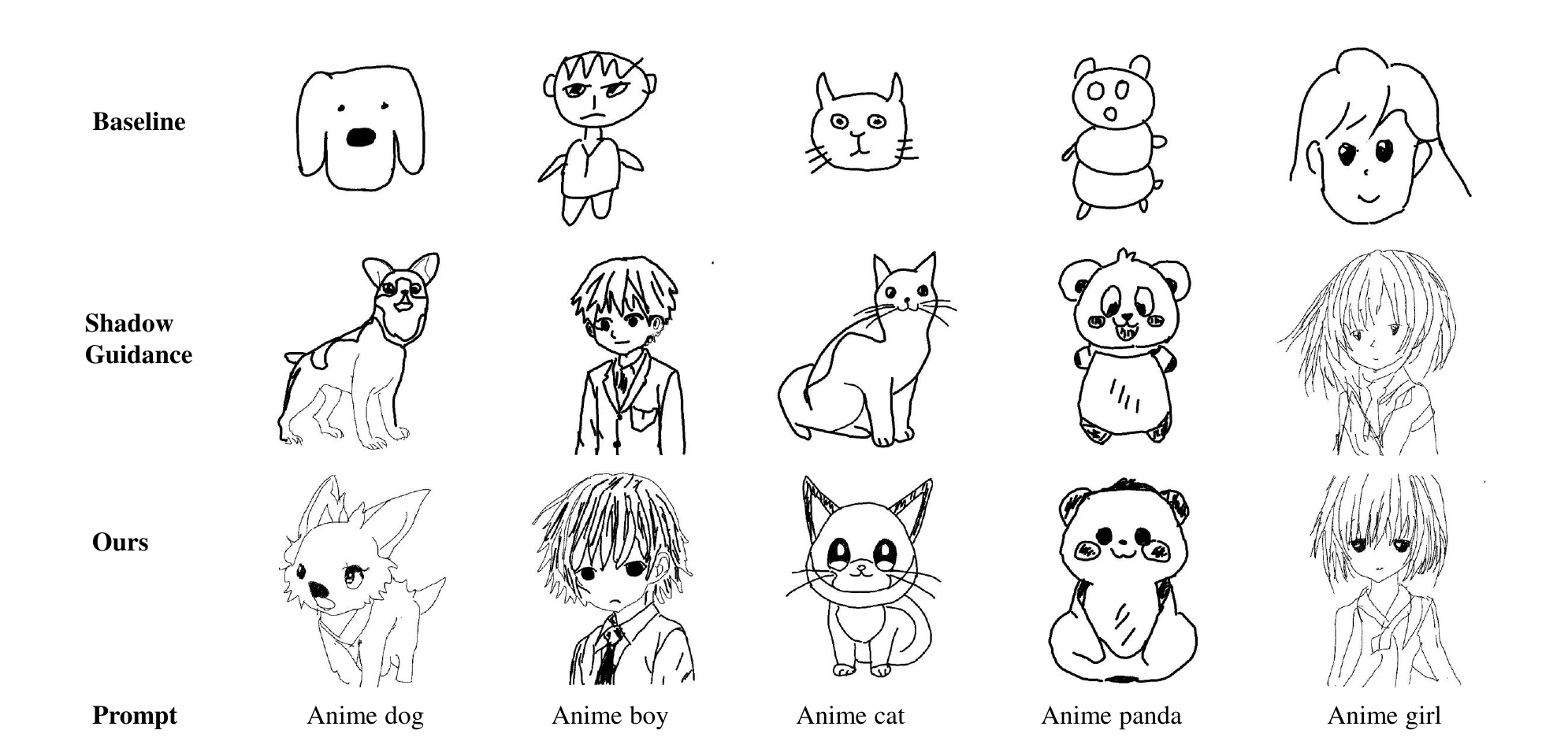}
    \caption{The results from the user study, showcasing drawings by users without drawing experience. The drawings exhibit an anime style, achieved using StreamDiffusion with the style LoRA.}
    \label{fig:results2}
\end{figure*}

\section{User Study}

We conducted a user study to evaluate the effectiveness of our proposed drawing assistance system and the impact of different guidance placements based on realistic style. We invited 8 participants (1 female and 7 males) to join this user study. Among them, 2 participants are experts, and 6 participants are novices. As one participant selected the same option for all questions in the questionnaire, we discarded his data. The final statistical results are based on the valid responses of 7 participants.

We set up three experimental groups for comparison: the baseline interface (Figure \ref{fig:ui(a)}), the shadow guidance interface (Figure \ref{fig:ui(b)}), and our guidance interface (Figure \ref{fig:interface}). Both the shadow guidance interface and our proposed interface update the guidance sketch in response to users' hand-drawn sketches. The shadow guidance interface places the guidance sketch at the bottom background layer of the canvas, while our interface places the guidance sketches on the right side of the canvas, allowing users to freely choose whether to display the guidance sketch in the background layer. Additionally, both the shadow guidance interface and ours are consistent in terms of image generation quality. Since the study involved three different drawing interfaces, we first conducted the control experiment using the interface without any drawing assistance. To avoid any order bias, the sequence in which the remaining two drawing assistance interfaces were used was randomized. All participants used a WACOM graphics tablet (512 × 512 drawing area) for portrait creation, operating on a single Windows computer equipped with a GeForce RTX 4090 GPU and an Intel Core i7-13700KF CPU.

Before the formal experiment, participants were asked to use all three drawing interfaces to familiarize themselves with the features. The experiment then presupposed five realistic categories. Participants were asked to choose three of them and to identify the subject they would be drawing about. Participants first used the baseline interface, followed by the shadow guidance and our interface in random order. For each interface, they created a drawing based on the same category. They were timed from the moment they began drawing until they finished. After completing the tasks, participants were asked to fill out a questionnaire assessing ease of use, drawing assistance effectiveness, and which interface helped improve their drawings. Based on the experiments conducted using the realistic style, we carried out the following evaluations. The questionnaire consisted of the System Usability Scale (SUS) and a user preference experiment. In addition, we invited 10 non-experts and 3 experts to evaluate the drawings produced by users with different interfaces.

\section{Results}
We analyzed the results from questionnaires and user drawing scores to evaluate the effectiveness of our drawing assistance system and to assess which type of assistance is more helpful in improving user drawings.

\subsection{User Experience Study}
 
Both the shadow guidance interface and our proposed guidance interface can infer the user's drawing intentions from incomplete sketches, generating high-quality guidance sketches. However, the two interfaces adopt different approaches to drawing assistance. The shadow guidance interface provides a single guidance sketch positioned at the bottom background layer of the canvas, whereas our interface offers four guidance sketches, allowing users the flexibility to choose whether to anchor them to the background layer.

Compared to the baseline interface, our guidance interface significantly enhances drawing quality, enabling users to create more detailed sketches. Table \ref{table:sus} presents the detailed scores from the System Usability Scale (SUS). Our guidance interface received an SUS score of 84.67, indicating high usability and effective drawing assistance. Among the evaluated questions, Question 5 received a score of 4.57, demonstrating that our system offers comprehensive drawing support. Other questions also received positive scores, further confirming that our drawing assistance system is recognized for its simplicity and ease of use. Figure \ref{fig:results1} shows some results produced using the baseline interface, the shadow guidance interface, and our proposed guidance interface, all generated results with realistic style. In addition, to demonstrate the effectiveness of our system in anime style illustration, we also conducted user study using the anime style. Figure \ref{fig:results2} showcases user-drawn sketches created under the anime style.

Additionally, we invited 10 non-experts and 3 experts to evaluate the drawings created using the three interfaces based on three evaluation metrics: overall shape, local proportion, and line quality. Non-experts are individuals without formal drawing experience, while experts are professionals with specialized expertise in drawing. The participants rated the drawings on a five-point scale, where 1 represents the lowest quality and 5 represents the highest quality. Given the difference in expertise between non-experts and experts in evaluation, we collected their results separately. 
Table \ref{table:drawing_scores} presents our evaluation scores for the five categories shown in Figure \ref{fig:results1}, based on the overall shape, local proportion, and line quality. Our system significantly improved user drawing results compared to the baseline interface, suggesting a positive influence on the drawing process. Note that both the shadow guidance interface and our proposed interface provided guidance sketches of equal quality, meaning any differences in final drawing scores do not directly indicate the superiority of one interface over the other. The user preference experiment comparing the shadow guidance interface and our proposed interface will be discussed further.

\begin{table}[htbp]
\caption{The result of the SUS questionnaire. $\Uparrow$ indicates that higher scores are better; $\Downarrow$ for the other case.} 
\label{table:sus}
\begin{center}     
\begin{tabular}{|c|>{\centering\arraybackslash}p{5.5cm}|c|c|}
\hline
 & Questions & Mean & SD  \\
\hline
    1 & \makecell[l]{I would like to use this system frequently. $\Uparrow$} & 4.43 & 0.73\\ \hline

    2 & \makecell[l]{I found this system unnecessarily complex. $\Downarrow$}  & 2.14 & 0.99\\ \hline
    3 & \makecell[l]{This system was easy to use. $\Uparrow$}  & 4.43 & 0.73\\ \hline
    4 & \makecell[l]{I would need the support of a technical person\\ to be able to use this system. $\Downarrow$} & 1.71 & 1.16 \\ \hline
    5 & \makecell[l]{I found the various functions in this system\\ were well integrated. $\Uparrow$} & 4.57 & 0.49\\ \hline
    6 & \makecell[l]{I thought there was too much inconsistency\\ in this system. $\Downarrow$} & 1.57 & 0.49\\ \hline
    7 & \makecell[l]{I would imagine that most people would learn\\ to use this system very quickly. $\Uparrow$} & 4.43 & 0.49\\ \hline
    8 & \makecell[l]{I found the system very cumbersome to use. $\Downarrow$} & 1.43 & 0.73\\ \hline
    9 & \makecell[l]{I felt very confident in using this system. $\Uparrow$} & 4.43 & 0.49\\ \hline
    10 & \makecell[l]{I needed to learn a lot of things before I could\\ get going with this system. $\Downarrow$} & 1.57 & 1.05\\ \hline
\end{tabular}

\end{center}
\end{table}

\begin{table}[htbp]
\centering
\caption{The drawing result scores for overall shape, local proportion, and line quality across the baseline interface, shadow guidance interface, and our interface for non-experts and experts.}
\label{table:drawing_scores}
\begin{tabular}{>{\centering\arraybackslash}p{1.3cm} >{\centering\arraybackslash}p{1.3cm} >{\centering\arraybackslash}p{1.3cm} >{\centering\arraybackslash}p{1.3cm} >{\centering\arraybackslash}p{1.3cm}}

\toprule
Methods & Group & Overall Shape & Local Proportion & Line Quality \\
\midrule
\multirow{2}{*}{Baseline} & Non-expert & 2.20 & 2.18 & 2.58 \\ \cline{2-5}
 & Expert & 2.40 & 2.40 & 2.70 \\
\midrule
\multirow{2}{*}{Shadow} & Non-expert & 3.07 & 3.11 & 3.40 \\ \cline{2-5}
 & Expert & 3.30 & 3.50 & 3.80 \\
\midrule
\multirow{2}{*}{Ours} & Non-expert & \textbf{4.40} & \textbf{4.36}  & \textbf{4.29}\\ \cline{2-5}
 & Expert & \textbf{4.10} & \textbf{4.20} & \textbf{4.40} \\
\bottomrule
\end{tabular}
\end{table}

\subsection{User Preference Experiment}

To investigate which drawing guidance method better helps users create drawings that align with their intentions, we designed a user preference experiment to evaluate the shadow guidance and our proposed interface. The eight participants in the user study completed the SUS questionnaire and then participated in our user preference experiment. We discarded the results of the participant who selected the same option for all responses. Table \ref{table:user performance} shows the distribution of user selections between the baseline interface, the shadow guidance interface, and our proposed interface for each question. In the survey, we asked users which interface best helped them achieve their initial drawing intentions. The results revealed that 85.7\% of users selected our interface, 14.3\% chose the shadow guidance interface, and none selected the baseline interface.

For the question \textit{``Which interface helped you achieve your desired outcome?"}, 71.4\% of users preferred our interface, 28.6\% chose the shadow guidance interface, and 0\% chose the baseline interface. Users who favored our proposed guidance interface provided feedback indicating that the fixed position of the guidance sketch in the shadow guidance interface hindered their ability to freely express their ideas, as the sketch remained fixed on the background layer. Regarding the question \textit{``Which interface do you think is better for your drawing?"}, 85.7\% of users selected our interface, while 14.3\% preferred the shadow guidance interface.

The results of the experiment demonstrate that our drawing guidance method makes it easier for users to create drawings that align with their expectations. By allowing users to keep the guidance sketch separate from the background layer during freehand drawing, our interface provides greater freedom to express creativity without the constraints of a fixed guidance sketch. This flexibility ultimately enhances the user’s drawing experience.

\begin{table}[htbp]
\caption{The result of user preference experiment.} 
\label{table:user performance}
\begin{center}     
\begin{tabular}{c c c c}
\toprule
\makecell[c]{Questions} & Baseline & Shadow & Ours \\
\midrule
\makecell[c]{Which experiment's results best\\ matched your initial expectations?} & 0 & 14.3\% & \textbf{85.7\%}  \\
\midrule
\makecell[c]{Which is your desired outcome\\ in the three groups?} & 0 & 28.6\% & \textbf{71.4\%} \\
\midrule
\makecell[c]{Which one do you think is better\\ for you to draw?} & 0 & 14.3\% & \textbf{85.7\%} \\
\bottomrule
\end{tabular}
\end{center}
\end{table}

\section{Conclusion}
This work presented a drawing guidance system that generated an anime illustration RGB image from a simple user's hand-drawn sketch and extracted sketch lines from the image to assist users in refining their drawings. The results from our user study demonstrated that the system effectively supported users with no drawing experience, enabling them to create sketches that aligned with their intended ideas. Additionally, the system proved beneficial for experienced users, assisting them in producing more creative drawings.

\begin{figure}[t]
    \centering
    \includegraphics[width=\linewidth]{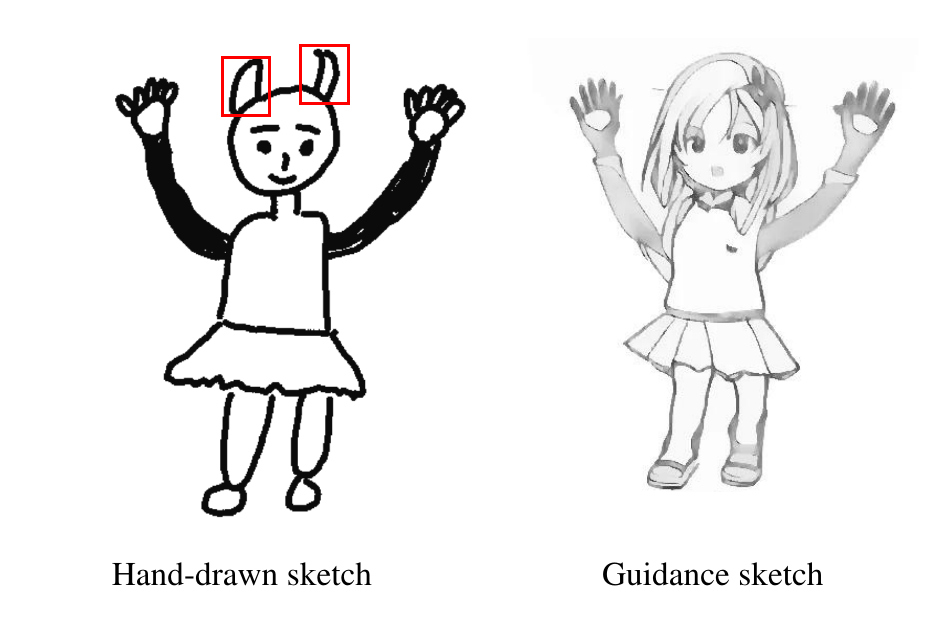}
    \caption{Example of failure case: the user drew horns on the character's head (left) but the generated guidance sketch may fail to generate horn parts (right). }

    \label{fig:buyizhi}
\end{figure}

Our system has several limitations and practical constraints. Although the introduced skip‑connection mechanism improves consistency between the input of hand-drawn sketch and the generated output, the generated images may still deviate from the user’s intent, as shown in Figure \ref{fig:buyizhi}. The provided prompt  was "anime, anime-style, a girl", and the sketch included a pair of horns on the character’s head. However, the generated guidance sketch failed to capture this feature, indicating an inconsistency in detailed feature generation. In addition, the current system has difficulty supporting local incremental updates, which may be solved using mask-based image generation approaches.

In future work, we plan to introduce a local incremental update mechanism that regenerates only the regions modified by the user, thereby preserving the consistency of unaltered areas. At the same time, due to the significant domain gap between line drawings and RGB images, the system may have limited capability in handling line drawings, which can lead to inconsistencies in the generated results. To address this, we will integrate conditional generation models such as ControlNet, which utilize the preprocessed line drawings as control signals to further enhance the alignment between the guidance sketch and the user’s hand-drawn sketch. In addition, it is worth systematically performing both quantitative and qualitative evaluations of diffusion models against GAN-based methods and traditional sketch retrieval techniques for real-time interactive drawing scenarios.

\section*{Acknowledgment}
This research result was based on the outcomes of the ``GENIAC (Generative AI Accelerator Challenge)'' project, implemented by the Ministry of Economy, Trade and Industry and the New Energy and Industrial Technology Development Organization (NEDO), aimed at strengthening domestic generative AI development capabilities. This research was supported by the Kayamori Foundation of Informational Science Advancement.

\bibliographystyle{ieeetr}
\bibliography{ref}
\end{document}